\documentclass[aps,twocolumn,showpacs]{revtex4}
\usepackage{graphicx}

\begin{document}
\title{Nanomechanical Analog of a Laser: Amplification of \\ Mechanical Oscillations by
Stimulated Zeeman Transitions}
\author{Igor Bargatin and M. L. Roukes}
\affiliation{Condensed Matter Physics, California Institute of Technology, MC
114-36, Pasadena, CA 91125} 
\begin{abstract}
We propose a magnetomechanical device that exhibits many properties of a
laser. The device is formed by a nanocantilever and dynamically polarized
paramagnetic nuclei of a solid sample in a strong external magnetic field. The
corresponding quantum oscillator and effective two-level systems are coupled
by the magnetostatic dipole-dipole interaction between a permanent magnet on
the cantilever tip and the magnetic moments of the spins, so that the entire
system is effectively described by the Jaynes--Cummings model. We consider the
possibility of observing transient and cw lasing in this system, and show how
these processes can be used to improve the sensitivity of magnetic resonance
force microscopy.
\end{abstract} \maketitle

The invention of masers and lasers in the middle of the twentieth century
\cite{firstlaser} has engendered whole new fields of science and myriads of
applications. This success of laser science and technology demonstrates the
value of the basic principles of laser devices and encourages one to look for
other systems in which these principles can be realized.

Regardless of the frequency range and other details of a practical
implementation, all laser-like devices involve one or more quantum-mechanical
oscillators resonantly interacting with a continuously pumped multilevel
quantum system. In the ubiquitous optical laser, the oscillator is realized by
a mode of a high-Q electromagnetic cavity, a mode resonant with optical
transitions of bound electrons in the active medium. Masers use microwave
transitions of gas molecules or electron spins of a paramagnetic solid in a
strong magnetic field. Finally, the active medium of a free-electron laser is
a relativistic electron beam, whose energy levels can be defined by a
specially configured magnetic field \cite{fel}.

This relative diversity of possible realizations of the active medium is not,
however, matched by the demonstrated realizations of the other essential part
of a lasing system, the oscillator. The authors are aware of only one
laser-like device that used an oscillator different from a field mode of an
electromagnetic cavity---the nuclear-magnetic-resonance (NMR) laser
\cite{nuclearchaos}. In that device, nuclear spins of a solid sample were
inductively coupled to a resonant LC circuit. Although many properties of an
LC circuit are strikingly different from those of a cavity resonator, one can
argue that the underlying physics in the two cases is the same: The
oscillations correspond to normal modes of a complex electromagnetic system,
whether it consists of an electromagnetic field confined by reflecting walls
or of coupled electric and magnetic fields of capacitive and inductive
elements.

In this Letter, we propose a laser-like device in which the oscillator is
realized by a fundamentally different kind of a device---a nanomechanical
resonator, e.g., a nanoscale cantilever or doubly clamped beam. Recent
advances in nanofabrication and detection techniques have pushed the
fundamental-mode frequencies of nanomechanical oscillators to the microwave
range \cite{henry_1GHz}, approaching the point where their properties begin to
be limited by quantum effects \cite{quantum}. In addition, micro- and
nanoelectromechanical oscillators generally exhibit low noise and high quality
factors, which naturally has led to applications for integrated high-frequency
signal generation and processing \cite{mems}. These properties make it
possible to use long-term coherent response of a high-frequency nanomechanical
oscillator in a laser-like device.

Similarly to the case of NMR laser and solid-state masers, we propose to use
nuclear or electron spins in a strong external magnetic field as the active
medium of a ``mechanical laser''. A nanomechanical oscillator can be
effectively coupled to the magnetic moments of such spins by incorporating a
small ferromagnetic tip on its surface. At microscopic distances between the
cantilever and the sample, the coupling between the tip and the spins is
essentially magnetostatic.

Systems consisting of a micro- or nanomechanical cantilever coupled to
resonant magnetic spins of a solid sample have been extensively studied in the
context of magnetic resonance force microscopy (MRFM) \cite{mrfm_rev}.
However, in all MRFM experiments performed so far, the fundamental frequency
of the cantilever was orders of magnitude below the Larmor frequency of the
magnetic spins. Resonant transfer of energy quanta from spins to the
mechanical oscillator is impossible in this case. Therefore, the resonance is
achieved by using an RF or microwave field to modulate the sample
magnetization at the cantilever frequency \cite{mrfm_rev}.

For the device proposed here, it is essential that the motion of the
mechanical oscillator be resonantly coupled to the free precession of magnetic
spins, which means that the frequency of the used mechanical mode must be
close to the Larmor frequency of the sample \cite{sidles92}. For conventional
experiments with external magnetic fields of a few Tesla, the Larmor
frequencies are of the order of tens of megahertz and tens of gigahertz for
nuclear and electron magnetic resonance, respectively \cite{slichter}. Given
that the highest fundamental-mode frequency of nanomechanical oscillators
measured so far is slightly above 1 GHz \cite{henry_1GHz}, resonant coupling
between mechanical oscillators and electron spins in strong magnetic fields
seems unfeasible at this time. Operation in low magnetic fields, on the other
hand, would prevent complete polarization of electrons and make the system
more sensitive to ambient magnetic fields. In the rest of this Letter, we will
therefore concentrate on the case involving nuclear spins.

Figure 1 shows a schematic of the proposed device. A nanomechanical
oscillator---a cantilever in this case---is positioned near a sample that
contains precessing nuclear spins, some of which are shown schematically in
the figure. A ferromagnet on the cantilever tip creates a magnetic field,
which can be approximated as the field of a magnetic dipole. When superposed
on the uniform external field ${\bf B_0}$, this field modifies the total
magnetic field seen by nuclear spins and, therefore, their Larmor frequency.
As a result, only a certain slice of the sample, known as the sensitive slice,
will have the Larmor frequency resonant with the frequency of the used mode of
the cantilever \cite{mrfm_rev}.

\begin{figure} \label{fig:schem}
\includegraphics[width=3in]{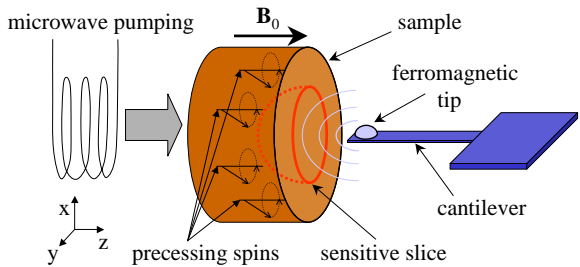}
\caption{Schematic of a mechanical laser device.}
\end{figure}

The rotating transverse component of nuclear magnetization couples to the
ferromagnetic tip via a dipolar magnetostatic interaction, with the resulting
force driving cantilever oscillations. Conversely, a moving ferromagnet
creates an AC magnetic field, oscillating at the frequency of the cantilever
motion, inside the sample. This RF field can drive transitions between Zeeman
levels of nuclear spins---stimulated transitions in the language of standard
rate-equation laser theory. The resulting coupled interaction---spins driving
the cantilever and the cantilever, in turn, driving the spins---leads to a
kind of positive feedback that arises in all laser-like systems.

Although a variety of nanomechanical devices could be used---involving, for
example, torsional or flexural modes---we focus here on nanocantilevers, which
are especially convenient for scanning with small tip-sample separations. The
device we propose can then be aptly termed a ``cantilaser''. To provide a
concrete example, we will assume the following parameters for the cantilever:
fundamental-mode frequency $\omega_c/2\pi=20$~MHz, effective spring constant
$k_c=0.1$ N/m, quality factor $Q=10^5$, the transverse magnetic field gradient
(due to the ferromagnetic tip) $\frac{\partial {\bf B}_\perp({\bf
r})}{\partial x}=1$~G/\AA $=10^6$~T/m within the sensitive slice. This
magnetic field gradient can be created by a rare-earth-metal magnet at a
distance of about 1 micron \cite{high_grad_MRFM}. At such relatively large
distances, one can usually neglect all nonmagnetic interactions between the
cantilever and the sample. Note also that the intrinsic Q factor of a
nanomechanical oscillator can be effectively increased by a few orders of
magnitude using positive feedback \cite{pos_feedback} or parametric pumping
\cite{darrel}.

For the parameters of the nuclear spin subsystem, we will take values
representative of crystalline materials \cite{slichter}: transverse relaxation
time $T_2=50\,\mu$s and nuclear gyromagnetic ratio $\gamma_n=2\pi\times
10$~MHz/T. The bulk of the sample material will be resonant with the
cantilever oscillations in an external field of $B_0=\omega_c/\gamma_n=2$~T.

In order to observe lasing in any system, one must introduce a pumping
mechanism to compensate for the energy dissipated in both the oscillator and
the active medium. For nuclear spins, such pumping can be produced by dynamic
nuclear polarization (DNP) \cite{slichter}. In this mechanism, microwave or
optical radiation is used to saturate an electron transition, causing them to
preferentially absorb photons of only one circular polarization. Some of the
absorbed angular momentum is then transferred from the electrons to the nuclei
of the sample through various equilibration processes. This technique has been
successfully employed to pump the NMR laser at the liquid helium temperature
(4.2~K) using a microwave source as in Fig.~1, with the effective pumping time
as low as $T_p= 0.2$ s \cite{nuclearchaos}. We will use this pumping rate and
the equilibrium longitudinal polarization $M_{eq}=-0.3$ that is achievable in
a 2-Tesla external field.

The dynamics of the cantilaser can be described by the Hamiltonian
$$\hat H=\hbar\omega_c \hat a^\dagger \hat a+
\hbar\gamma_n \sum_i {\bf B}_i \cdot\hat {\bf S}_i +\hbar(\hat a^\dagger+\hat
a) \sum_i 2{\bf g}_i\cdot\hat {\bf S}_i+\hat H_r,$$ where $a^\dagger$ and $a$
are the creation and annihilation operators of the cantilever mode, $\hat {\bf
S}_i$ is the spin operator of $i$th nucleus, $\hat {\bf B}_i$ is the the
external field at the site of the $i$th spin, ${\bf
g}_i=\frac{\gamma_n}{2}\frac{\partial {\bf B}({\bf r}_i)}{\partial
x}\sqrt{\frac{\hbar\omega_c}{2k_c}}$ is the vector constant of the coupling
between the $i$th spin and the cantilever, and $\hat H_r$ describes
relaxation-inducing couplings to the environment. This Hamiltonian was first
considered by Jaynes and Cummings, who used it describe quantum behavior of
masers \cite{jcmodel}. In the same paper, they also showed that the
corresponding dynamics can usually be described by semiclassical equations,
which treat the resonator classically and the spins quantum mechanically.

Considering the nuclear spins in their respective rotating frames (as defined
by the local field ${\bf B}_i$ and field gradient $\frac{\partial {\bf B}({\bf
r}_i)}{\partial x}$ \cite{sidles92}) and using the slowly-varying-amplitude
approximation for the cantilever, we can write such semiclassical equations in
the form
\begin{equation}\label{Bloch}
 \begin{array}{l}
 \dot{A}+\kappa A = -gNM_-,\\
 \dot{M_-}+\Gamma_\perp M_-=gM_z A,\\
 \dot{M_z}+\Gamma_\| (M_z-M_{eq})=-g(M_- A^*+M_-^* A)/2,
\end{array}\end{equation}
where $A$ is the (generally complex) amplitude of cantilever oscillations,
normalized by the amplitude, $\sqrt{\hbar\omega_c/(2k_c)}\approx 2.6\cdot
10^{-13}$ m, of zero-energy quantum motion, $\kappa=\omega_c/2Q\approx
630\,{\rm s}^{-1}$ is the cantilever decay rate, $N$ is the number of resonant
spins in the sensitive slice, $M_-$ and $M_z$ are the normalized
($|M_-|^2+M_z^2\leq 1$) transverse and longitudinal (with respect to ${\bf
B_0}$) components of nuclear polarization, $\Gamma_\perp=T_2^{-1}= 20\cdot
10^3\,{\rm s}^{-1}$ and $\Gamma_\|=T_p^{-1}= 5\,{\rm s}^{-1}$ are the
effective transverse and longitudinal polarization relaxation rates, and
$g=\frac{\gamma_n}{2}\left|\frac{\partial {\bf B}_\perp({\bf r})}{\partial
z}\right|\sqrt{\frac{\hbar\omega_c}{2k_c}}\approx 8.0\,{\rm s}^{-1}$ is the
scalar coupling constant of the interaction between the cantilever and the
nuclear spins.

In equations~(\ref{Bloch}) we implicitly assumed that all resonant nuclei in
the sensitive slice are spin-half and that they all see the same strength and
gradient of the magnetic field. The latter is an obvious simplification since
in MRFM experiments, the magnetic-resonance frequency and coupling strength
varies continuously over the sensitive slice \cite{suter}. However, the same
problem of inhomogeneous broadening and nonuniform coupling arises in most
quantum optics and laser setups \cite{haken}, and it was found experimentally
\cite{haroche83,kimble95} that equations of the form (\ref{Bloch}) still
correctly reproduce most features of the coupled spin--oscillator dynamics. In
this Letter, we will therefore restrict our analysis to the simplest model of
Eqs.~(\ref{Bloch})

It is easy to find the steady-state solutions of Eqs.~(\ref{Bloch}). The
nontrivial lasing solution may exist only in the case of population inversion,
$M_{eq}<0$, and is given by $A_{cw}=\sqrt{(N|M_{eq}|-N_t)\Gamma_\|/\kappa},$
where $N_t=\kappa\Gamma_\perp/g^2$ is the threshold population inversion.
Substituting $M_{eq}=-0.3$ and other parameter values given above, we find
that in order to support cw lasing, the number of atoms in the sensitive slice
should be $N>N_{cw}=N_t/|M_{eq}|\approx 0.65\cdot 10^6$. This may seem like a
large number; however, even an atomically thin sensitive slice of a
homogeneous sample contains of the order of $10^7$ nuclei if the diameter of
the sensitive slice is just $1\, \mu$m. Much larger sensitive slices have been
used in nuclear MRFM experiments so far, so exceeding the lasing threshold
seems quite feasible.

One of the more interesting transient phenomena predicted by the
Jaynes--Cummings model is the coherent oscillation of population between the
oscillator and spins \cite{jcmodel}, an effect similar to the oscillations of
energy between two weakly coupled classical harmonic oscillators. Also known
as ringing superradiance, this phenomenon has been observed in different
quantum-optical systems \cite{haroche83,kimble95}. In order for the energy
oscillations to be observable in a cantilaser, the effective frequency of the
oscillations, equal to $\sqrt{|M_{eq}|N}g$ \cite{jcmodel}, should be larger
than the fastest relaxation rate of the system, $\Gamma_\perp$. We can
therefore roughly estimate the minimum number of atoms necessary to observe
the oscillations as $N_{sr}=\Gamma_\perp^2/(|M_{eq}|g^2)\approx 20\cdot 10^6$.
As we show below, our numerical simulations support the validity of this
estimate.

Another interesting transient predicted by the Jaynes-Cummings model is a
solitary pulse that irreversibly depletes the energy stored in the active
medium. Known as giant pulses in the standard laser theory \cite{haken}, these
transients can appear if $\Gamma_\perp> |M_{eq}|Ng^2/\Gamma_\perp>\kappa$,
which implies $N_{sr}>N>N_{cw}$. Such a giant pulse reduces the population
inversion to zero and therefore consumes one half of the total potential
energy of the active medium (i.e., the sensitive slice, which can in principle
be adjusted to encompass most of the sample \cite{suter}). This is in contrast
to the case of ringing superradiance, where all of the available energy
oscillates back and forth between the cantilever and spins.

\begin{figure}\label{fig:tran}
\includegraphics[width=3in]{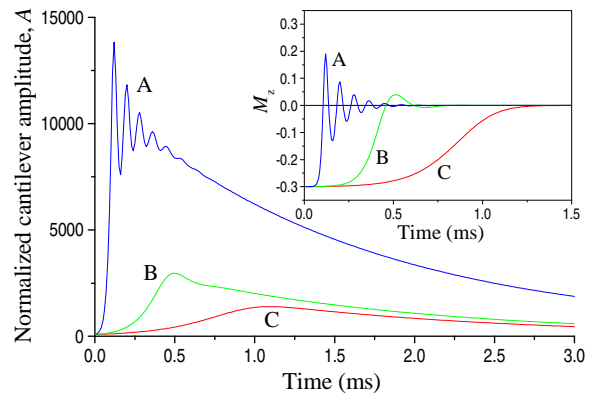}
\caption{Characteristic transients of a cantilaser for different number of
resonant nuclei within the sensitive slice: (A) $N=200\cdot10^6$ , (B)
$15\cdot 10^6$, and (C) $5\cdot 10^6$. The main panel shows the normalized
amplitude of cantilever oscillations, $A(t)$, the inset shows the longitudinal
nuclear polarization, $M_z(t)$. The initial conditions are
$A(0)=94,\,M_-(0)=0,\,M_z(0)=-0.3$.}
\end{figure}

Figure 2 shows three characteristic transient outputs of a cantilaser,
obtained by numerical integration of Eqs.~(\ref{Bloch}). As the number of
resonant nuclei in the sensitive slice decreases from $N=200\cdot10^6\gg
N_{sr}$ to $N=5\cdot 10^6\ll N_{sr}$, the frequency and amplitude of energy
oscillations decreases until just one ``giant'' pulse is observed. If one
further keeps decreasing the number of atoms, the single pulse becomes longer
and smaller in amplitude until it disappears completely as the number of atoms
goes below the cw lasing threshold $N_{cw}$. Note that the tails of the output
transients always decay at the time scale of $\kappa^{-1}$ because cantilever
decay is the dominant mechanism of energy dissipation here. In contrast, the
coherent oscillations, if present at all, decay at the time scale of
$\Gamma_\perp^{-1}=T_2$ since spin-spin relaxation is the dominant mechanism
for the loss of coherence.

The initial conditions for pulsed transients used in the simulations of Fig.~2
can be produced by a Q-switching technique \cite{haken}. Note that the initial
nuclear polarization, $M_-(0)=0,\,M_z(0)=M_{eq}$, is simply the equilibrium
polarization achieved in the presence of dynamic nuclear pumping and
negligible interaction with the cantilever. Also, the initial cantilever
amplitude corresponds to thermal vibrations of the cantilever at the
temperature $T=4.2$ K: $A(0)=A_{th}=\sqrt{2k_BT/(\hbar \omega_c)}\approx 94$,
where $k_B$ is the Boltzmann constant.

Since a cantilaser shares so much in its design and principles of operation
with MRFM setups, it is natural to consider whether the effects describe above
can be used to improve the sensitivity of nuclear MRFM. The single-shot
sensitivity of the first nuclear MRFM experiment \cite{rugar94} was
approximately $10^{13}$ thermally polarized nuclear spins at room temperature
or about $10^{11}$ nuclear spins at 4.2~K (nuclear polarizability is inversely
proportional to temperature). Since then, the sensitivity of MRFM experiments
has been improved to about 100 Bohr magnetons \cite{100spins}, a magnetic
moment that is created by roughly $10^8$ nuclear spins at 4.2~K. The closest
competing technology---scanning SQUID-based magnetometry---has so far
demonstrated a sensitivity of $10^5$ Bohr magnetons \cite{squidNMR}, or
$10^{11}$ nuclear spins at 4.2~K.

To consider the MRFM sensitivity of a cantilaser, we will calculate the ratio
of power spectral density of lasing outputs to the power spectral density of
thermomechanical noise. Since the resonant frequency of the cantilever and the
Zeeman transition frequency of spins coincide, we can express all energy
quantities in terms of the number of energy quanta $\hbar\omega_c$. The power
of the thermomechanical noise is then proportional to
$n_{th}=A_{th}^2/2=k_BT/(\hbar\omega_c)\approx 4400$, and its bandwidth is
$\Delta\omega_{th}=\kappa$. Well above the lasing threshold, the power of the
cw lasing signal is
$n_{cw}=A_{cw}^2/2\approx\Gamma_\|N|M_{eq}|/(2\kappa)\approx N/840$, and its
bandwidth is $\Delta\omega_{cw}\approx\kappa (n_{th}+1/2)/n_{cw}$
\cite{haken}. The signal-to-noise ratio for the cw output is then
$SNR_{cw}=\frac{n_{cw}/\Delta\omega_{cw}}{n_{th}/\Delta\omega_{th}}\approx\left(n_{cw}/n_{th}\right)^2
\approx \left(N/3.7\cdot 10^6\right)^2$. The quadratic increase in $SNR_{cw}$
with the number of atoms reflects the spectral narrowing of the cw signal at
high output power, a fact that is well known and used in optical lasers
\cite{haken}.

For both kinds of pulsed outputs we considered (ringing superradiance and
single pulses), the efficiency of the energy transfer from spins to the
cantilever mode is of the order of unity. The peak cantilever amplitude of a
pulse in both cases then corresponds to mode population of about $n_{pulse}=N
|M_{eq}|/2$. Since all pulsed outputs eventually decay at the time scale of
$\kappa^{-1}$, their bandwidth can be taken to be
$\Delta\omega_{pulse}\approx\kappa$. Proceeding as above, we find
$SNR_{pulse}=N |M_{eq}|/(2 n_{th})\approx N/29000$. A cantilaser operating in
the pulsed mode would therefore have a single-shot sensitivity of about
$3\cdot10^4$ nuclear spins at 4.2~K, which is at least three orders of
magnitude better than the sensitivity of any existing alternative. A large
part of this improvement derives from the hyperpolarization of nuclei by DNP
processes, but the near-unity efficiency of energy transfer between spins and
cantilever in pulsed transients is also significant.

We conclude by considering different possible perspectives upon mechanical
lasing---from the standpoints of quantum optics, NMR spectroscopy, and MRFM.
From the perspective of quantum optics, the cantilaser is very similar to a
cavity QED system \cite{cqed}, albeit one with weak but comparable coupling
strength and longitudinal relaxation, $g\sim\Gamma_\|\ll
(\Gamma_\perp,\kappa)$, and high thermal population, $n_{th}\gg 1$. Since such
combinations of parameters are not available in quantum optical systems or NMR
laser, this opens up new possibilities for studying coherent quantum phenomena
in coupled oscillator--atom systems.

In conventional NMR studies, the possibility of a positive feedback between
the sample and the detecting resonance circuit has been long recognized.
Bloembergen first considered the back action of the detecting coil on the
sample almost 50 years ago \cite{bloembergen54}. Unfortunately, such back
action tends to shorten the signal pulses and therefore broaden spectral
features. This explains why this positive-feedback effect, known to NMR
practitioners as radiative damping, is generally undesirable in
high-resolution NMR spectroscopy. In contrast, MRFM experimentalists are not
interested in fine details of NMR spectra. Since the ultimate goal of MRFM is
atom-by-atom 3D mapping of nanoscopic objects, the required spectral
resolution should only be sufficient to distinguish between different nuclear
species. MRFM practitioners are therefore willing to trade fine spectral
resolution for signal strength and spatial resolution. This is exactly what a
mechanical laser provides, by embracing and fully exploiting the positive
feedback in the coupled oscillator--spin system.

We are grateful for support of this work from DARPA DSO/MOSAIC under ONR grant
N00014-02-1-0602, and from the NSF under grant ECS-0116776. We also thank Ed
Myers and John Sidles for helpful comments.

\end{document}